# High-efficiency narrowband terahertz generation in novel wide-bandgap barium chalcogenide crystals

D. Z. Suleimanova,[1] E. A. Migal,[1] D. V. Badikov,[2] and F. V. Potemkin[1,a)]

[1] *Faculty of Physics M.V. Lomonosov Moscow State University, Moscow, 119991, Russia*

[2] *High Technologies Laboratory, Kuban State University, Krasnodar, 350040, Russia*

We report on extensive experimental investigation of novel barium chalcogenides ($BaGa_4Se_7$, $BaGa_2GeSe_6$ and $BaGa_2GeS_6$) for generation of a strong narrowband terahertz (THz) radiation. The Ba-containing compounds demonstrate the large bandgap values, high optical damage threshold and wide transparency range from the visible to mid-infrared region, coupled with narrowband transparency windows in the THz frequency range. In these windows a THz generation via the second-order nonlinear process of optical rectification occurs under optical pumping by a 50-GW high-power Cr:forsterite laser leading to intrinsically narrowband THz pulses. The highest optical-to-THz conversion efficiency of $0.8 \times 10^{-5}$ is reached in the Z-cut $BaGa_2GeS_6$ crystal in the saturation regime and corresponds to an output energy of 20 nJ. Ba compounds provide narrow-bandwidth THz pulses at the frequencies in the range from 1.9 to 2.5 THz with a relative bandwidth of 3-4%, necessary for imaging and spectroscopy for security or medical applications.

Narrowband terahertz (THz) radiation, covering 0.1 to 10 THz frequency range, serves as an advantageous tool for both practical and scientific applications including biomedical imaging [1], THz control of quantum materials [2], nonlinear phononics [3] and THz photonics of 2D materials [4]. Narrowband THz sources, characterized by bandwidths under 0.1 THz, are particularly valuable for improvement of the spectral resolution in THz spectroscopy, allowing precise detection of molecular fingerprints, and for the resonant control and manipulation over matter. Nevertheless, achieving efficient narrowband THz generation remains challenging.

Conventional nonlinear nonorganic (ZnTe, $LiNbO_3$) and organic (DAST, DSTMS, OH-1) crystals have already been extensively used for broadband THz generation via optical rectification [5]. While narrowband THz radiation from these materials require complex chirped pulses difference frequency mixing schemes or employ quasi-phase matched periodically poled nonlinear crystals with a specially engineered multi-layer structure. In particular, a chirp-and-delay set-up allowed the

---

[a)] Electronic mail: potemkin@physics.msu.ru

generation of narrowband, multicycle THz pulses at 1.8 THz with a relative bandwidth up to 2.5% [6-9].

On the contrary, when exploring novel THz materials, the monoclinic BaGa$_4$Se$_7$ crystal demonstrated narrowband THz generation at frequencies of 1.97 and 2.34 THz under optical rectification, that was associated with the dense phonon mode distribution of the crystal [10]. The optical rectification experiments were carried out using Ti:sapphire oscillator, delivering pulses at a repetition rate of 5.1 MHz with an average power of 25 mW. Thus, the resulted optical-to-THz conversion efficiency achieved rather low values of 6.3×10$^{-8}$ [11,12]. More recently the monoclinic semi-organic crystal GUHP under cryogenic cooling demonstrated narrowband THz generation at frequencies of 1.1 and 1.67 THz with a relative bandwidth of 1%. However, the energy of THz pulses reached only 700 pJ at low temperatures (10 K), which corresponds to the conversion efficiency of 8×10$^{−7}$ [13]. Table 1 summarizes the recent achievements in the development of narrowband THz sources. Other materials such as organic molecular ferroelectric DCMBI [14], inorganic InSb [15], lithium ternary chalcopyrites [16] also demonstrate narrowband THz emission in the vicinity of phonon modes associated with the impulsive stimulated Raman scattering, which was mostly exploited for investigation of crystals' structure and optical properties rather than alternative sources of narrowband THz emitters. Nevertheless, these studies suggest the interest for the development of novel materials capable of narrowband THz radiation generation.

TABLE I. Parameters of narrowband THz sources based on semiconductor and organic materials.

| Nonlinear crystal | Optical-to-THz conversion efficiency (%) | Central frequency, THz (bandwidth, %) | Method | Reference |
|---|---|---|---|---|
| PPLN | 1x10$^{-3}$ | 0.5 (3.3%) | Chirp and delay | [6] |
| OH1 | 2x10$^{-4}$ | 1.1 (2.7%) | Chirp and delay | [7] |
| DSTMS | 2.5x10$^{-4}$ | 3.87 (2%) | Chirp and delay | [8] |
| HQM-TMS | 3.5x10$^{-5}$ | 0.54 | Chirp and delay | [9] |
| GUHP | 8×10$^{−7}$ | 1.1 and 1.67 (1%) | Direct OR | [13] |
|  | 6.3x10$^{-8}$ | 1.97 and 2.34 (2.5%) | Direct OR | [11,12] |
| BaGa$_4$Se$_7$ (BGSe) | 0.4×10$^{-5}$ | 1.9 (3.2%) | Direct OR | This work |
| BaGa$_2$GeSe$_6$ (BGGSe) | 0.54×10$^{-5}$ | 2.05 (3.7%) | Direct OR | This work |
| BaGa$_2$GeS$_6$ (BGGS) | 0.8×10$^{-5}$ | 2.54 (3.1%) | Direct OR | This work |

Quaternary barium sulfides and selenides have appeared to be promising materials for highly efficient generation of mid-IR pulses and have attracted significant attention from researchers due to their unique properties such as high nonlinearity, high damage threshold and chemical stability of the polished surface in air [17-19]. Successful growth of large size high optical quality BGGS and BGGSe crystals by the vertical Bridgman–Stockbarger technique was reported in [20]. However, the potential of these chalcogenides for THz applications remained unexplored, offering a novel avenue for investigation.

In this Letter, we report on the first experimental observation of strong narrowband THz generation in novel quaternary BaGa$_2$GeS$_6$ and BaGa$_2$GeSe$_6$ crystals via the second-order nonlinear process of optical rectification when exciting by Cr:forsterite laser. We compare these crystals with BaGa$_4$Se$_7$ in terms of output spectrum and conversion efficiency. Narrowband (with relative bandwidth of 3-4%) radiation was achieved in the frequency range from 1.9 to 2.5 THz. The highest optical-to-THz conversion efficiency of $0.8\times10^{-5}$ and corresponding output energy of 20 nJ were achieved in the Z-cut BaGa$_2$GeS$_6$ crystal in the saturation regime. The results reveal the potential of these quaternary Ba compounds for narrowband THz generation, paving the way for applications in biomedical imaging and narrowband spectroscopy.

Optical rectification experiments were performed using 500 µm thick Z-cut quaternary BaGa$_2$GeS$_6$ (BGGS), BaGa$_2$GeSe$_6$ (BGGSe) and ternary BaGa$_4$Se$_7$ (BGSe) crystals (Fig. 1 (b-c)). Here, crystallo-physical Z axis coincides with c-crystallographic axis for all studied crystals and denotes for the normal to the plane of the cut. For trigonal BGGSe and BGGS crystals it is normal to the $(a-b)$ plane, while for monoclinic BGSe crystal it is normal to (X-Y) plane, containing X and Y crystallo-physical directions [21]. In particular, Y crystallo-physical direction coincides with $b$-crystallographic axis and is orthogonal to X crystallo-physical direction.

High-power Cr:forsterite laser system delivering 100-fs pulses centered at 1.24 µm served as a pump source [22]. The experimental setup for the THz radiation measurements is shown in Fig. 1.

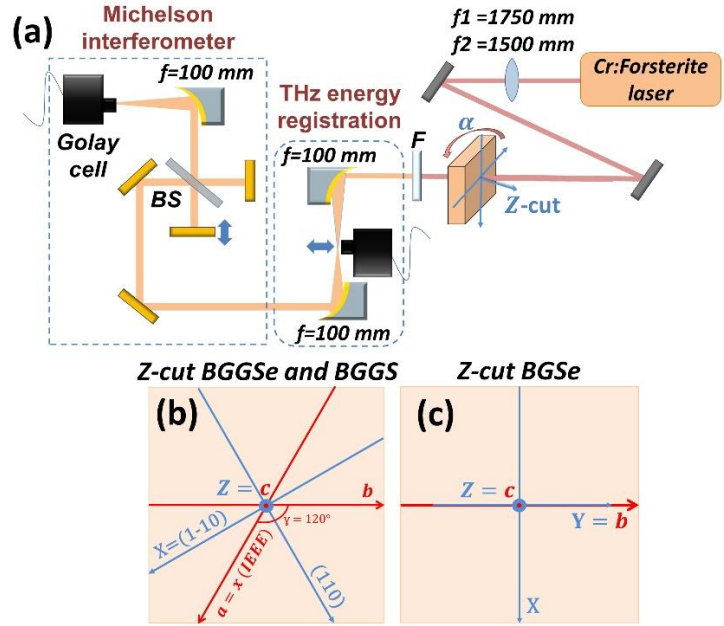

FIG. 1. (a) Experimental setup used to observe narrowband THz generation in barium crystals. $\vec{E}_{THz}$ represents the generated THz electric field, $\vec{E}_{pump}$ represents the electric field polarization of the Cr:Forsterite pump pulse. BS, beam splitter; F, low-pass THz filters, α, the crystal's rotation angle. (b-c) Planes of the cut for trigonal BGGSe, BGGS (b) and monoclinic BGSe (c) crystals with indication of crystallo-physical axes used in this work and the conventional crystallographic axes (abc).

To achieve efficient THz generation under optical rectification the pump beam with energy of 3.5 mJ was focused to a diameter of 6 mm (measured at the level of $1/e^2$) using a lens ($f_1$) with a focal length of 1.75 m, so that pump pulse fluence reached 10 mJ/cm$^2$ at the crystal surface (taking into account the Fresnel losses of crystals). The pump beam had a converging wavefront at the crystal position. All studied crystals were rotated around the propagation direction of the incident pump pulses in order to determine the optimal orientation of each crystal. The crystal's rotation angle is expressed by α, which is determined as the angle between the pump polarization direction and the Y crystallo-physical direction for BGSe and $a$-crystallographic axis for quaternary BGGS and BGGSe. Note that pump polarization was maintained to be linear and horizontal in the conducted experiments. The generated THz radiation energy was measured using a commercial calibrated Golay cell detector (Tydex, GC-1P) and was separated from the residual near-infrared pump radiation using low-pass THz filters (Tydex LPF10.9-24 and Tydex LPF23.4).

Figure 2(a) shows the dependence of the THz radiation energy on the rotation angle α. As a result, when using the BGGSe and BGGS crystals, the highest THz radiation energy was obtained, when either one of the crystallographic axes, or the 110 direction was set parallel to pump polarization direction ($\vec{E}_{pump}||a$, $\vec{E}_{pump}||b$ or $\vec{E}_{pump}||110$). Since both quaternary chalcogenides belong to the trigonal crystal system (point group 3) [18], the dependence of the generated THz radiation energy on the crystals' orientation appeared to be the same for both compounds. BGSe crystal, belonging to point group $m$ (monoclinic system), demonstrates the different behavior of the output THz energy.

In particular, the output THz energy reached 13 nJ in the 0.5 mm-thick Z-cut BGGSe crystal. This corresponds to an optical-to-THz conversion efficiency of $4.6 \times 10^{-6}$. As for the sulfide BGGS compound, the THz radiation energy of 8 nJ was achieved, that corresponds to the lower $2.7 \times 10^{-6}$ conversion efficiency. While for the reference ternary BGSe crystal, the maximum optical-to-THz conversion efficiency was obtained, when the pump polarization direction coincided with one of its crystallo-physical axes (Fig. 2(a)). It reached $3.8 \times 10^{-6}$ value and exceeded the previously obtained one in [11].

Spectra of generated THz pulses were characterized by a home-built Michelson-type interferometer combined with the same Golay cell detector as depicted in Fig. 1. A high-resistance Si-plate (Tydex HRFZ-Si) was used as a beam splitter, that provides a transmission of ~55 % over a wide frequency range up to 10 THz. The spectral resolution of the order of 13 GHz was achieved. A delay scan in the interferometer provides THz field first order autocorrelation functions, shown in Fig. 3(a), from which the spectral power was obtained by the Fourier transform.

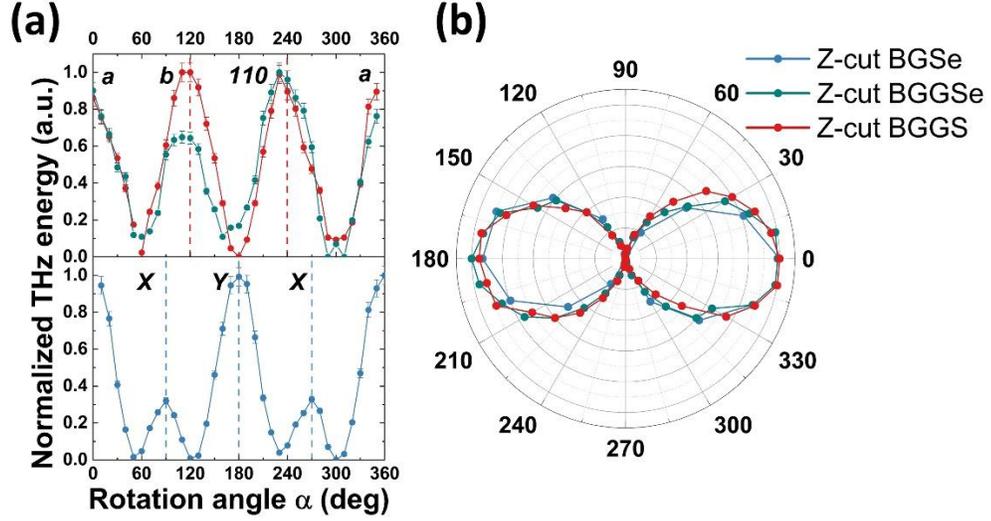

FIG. 2. (a) The dependence of the THz radiation energy on the crystal's rotation angle α. Different colors indicate for different crystals used (BGGS, BGGSe and BGSe). (b) Polarization of generated THz radiation measured in the experiment. The vertical and horizontal axes correspond to vertical and horizontal polarizations, respectively. For BGGSe and BGGS crystals pump polarization was set parallel to a-crystallographic axis, while for BGSe crystal – to Y crystallo-physical axis.

The emission obtained from investigated barium crystals is a multi-cycle THz pulse with the corresponding intensity spectrum shown in Fig. 3(b). As a result, narrowband THz radiation was achieved in the frequency range from 1.9 THz to 2.5 THz. For quaternary chalcogenides narrow THz radiation peaks are centered at frequencies of 2.05 THz and 2.55 THz with a full-width at half-maximum (FWHM) of 75 GHz ($\Delta f/f = 3.7\%$) for BGGSe and 80 GHz ($\Delta f/f = 3.1\%$) for BGGS, respectively. Note that sulfide compound also shows a weak frequency component at 2.05 THz. As for the reference BGSe crystal THz generation at 1.9 THz was obtained with a relative bandwidth of 3.2%, which is in a good agreement with the results achieved earlier [11,12].

To investigate the nature of such a narrowband THz radiation generation, observed in the experiment, the transmittance spectra of all studied crystals were measured (Fig. 3(b)). For this aim organic 500 µm-thick DAST crystal served as a source of broadband (1 – 4 THz) THz radiation. Specific narrowband transparency windows are observed across the measured transmittance spectra of barium chalcogenides, that are known to be influenced by numerous phonon modes [10,23,24]. Moreover, spectra of generated THz radiation appeared to lie within these narrow transparency gaps. Therefore, we attribute such a narrowband THz radiation generation to the optical rectification process that occur in the narrowband transparency windows within the limited bandwidth of the 100-fs pump pulse.

Polarization of generated THz radiation was characterized with the wire grid THz polarizer (Tydex) and appeared to be linear, which is crucial for further applications of THz radiation (Fig. 2(b)).

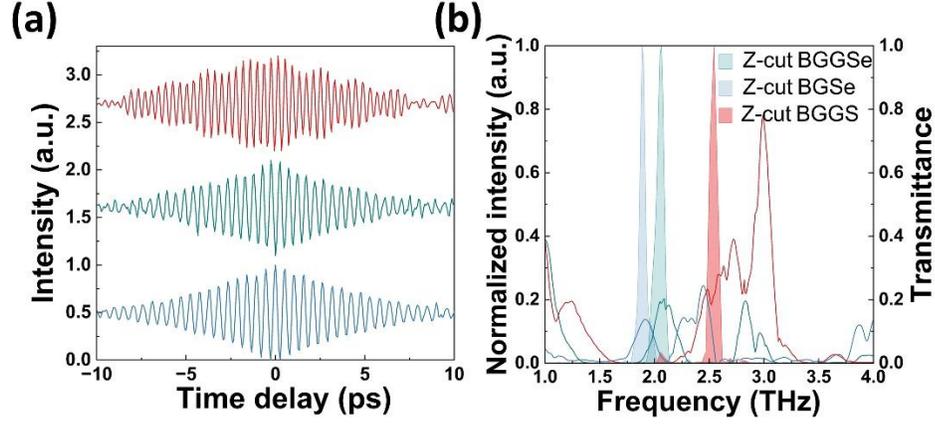

FIG. 3. (a) Measured THz field autocorrelations, (b) generated THz radiation (color-filled) and transmittance THz spectra (lines) of Z-cut BGSe, BGGSe and BGGS crystals. Different colors indicate for different crystals.

Thus, the strongest THz radiation obtained in quaternary trigonal compounds appeared to be polarized along $a$-crystallographic axis and corresponded to the pump polarization direction. Such dependence on excitation polarization and crystal orientation is governed by the second-order nonlinear polarization tensor and depends significantly on the values of non-zero nonlinear coefficients for a particular crystal.

Finally, in order to understand the possibilities of further increase the optical-to-THz conversion efficiency, the THz radiation energy dependence on the pump fluence for studied Z-cut BGGSe, BGGS and BGSe crystals was investigated (Fig. 4). For this aim, pump beam was focused to a smaller diameter of 2.4 mm (measured at the level of $1/e^2$) using another lens ($f_2$) with a focal length of 1.5 m, so that pump fluence reached higher values of up to 60 mJ/cm$^2$ at the crystal surface (taking into account the Fresnel losses of crystals). Then the pump pulse energy was varied in the experiment.

Since sulfide compound possesses the larger bandgap (3.37 eV [17]) compared to its selenide counterparts (2.73 eV and 2.38 eV for BGSe and BGGSe, respectively [17]), the higher values of pump intensity were applied. For instance, the pump intensity of the order of 155 GW/cm$^2$ was safely maintained in BGGS crystal when using the 100-fs 1.24-µm pump pulses and no damage threshold of the crystal surface occurred [19].

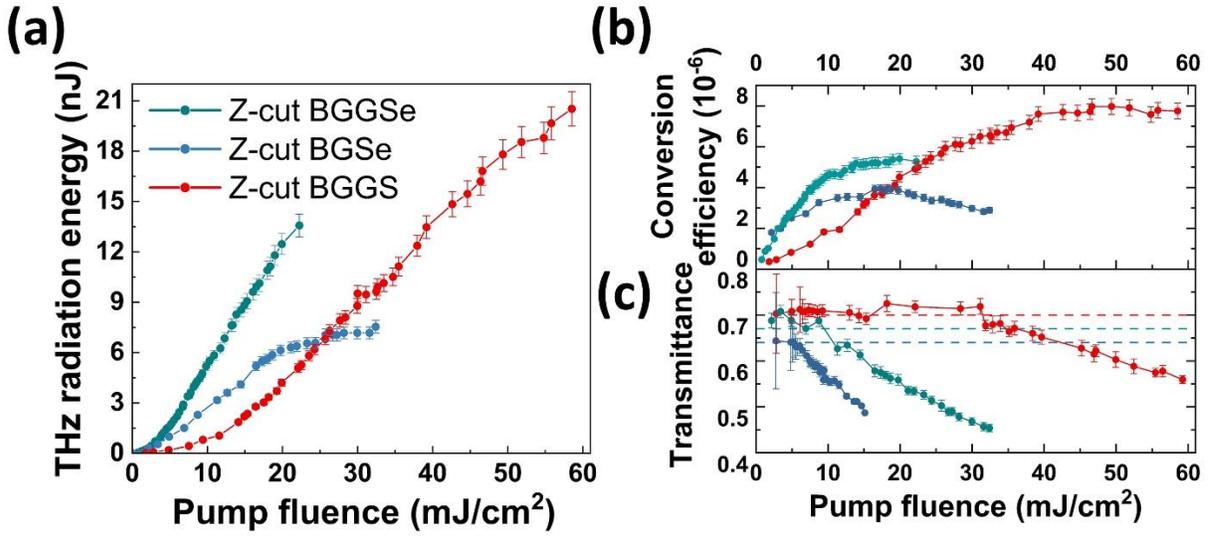

FIG. 4. (a) The THz radiation energy, (b) the corresponding optical-to-terahertz conversion efficiency and (c) the pump energy transmittance dependence on the pump pulse fluence for Z-cut $BaGa_2GeSe_6$ (BGGSe), $BaGa_2GeS_6$ (BGGS) and $BaGa_4Se_7$ (BGSe) crystals. Dashed lines correspond to the Fresnel losses of crystals. The pump pulse fluence was calculated taking into account the Fresnel losses of crystals. Different colors indicate for different crystals.

As the pump fluence was increased, the saturation of conversion efficiency became obvious for all investigated crystals. We attributed saturation to the appearance of multiphoton absorption since nonlinear losses were observed at high pumping level as shown in Fig. 4 (c). As a result, the highest optical-to-THz conversion efficiency of $7.9\times10^{-6}$ was achieved in Z-cut BGGS crystal in saturation regime and corresponding THz energy of 20 nJ was obtained. Note that saturation of THz generation in barium selenides was observed when pump fluence reached lower values of 15 mJ/cm$^2$, that can be attributed to the lower bandgaps of these crystals. Thus, we limited the applied pump pulse intensity (fluence) to the values of 200 GW/cm$^2$ (20 mJ/cm$^2$) and 300 GW/cm$^2$ (30 mJ/cm$^2$) for BGGSe and BGSe crystals due to the observed saturation of the conversion efficiency. As a result, in highly saturated regime the conversion efficiency of $5.4\times10^{-6}$ and $3.9\times10^{-6}$ was achieved for BGGSe and BGSe, respectively. While for the sulfide compound the pump intensity of 600 GW/cm$^2$ was reached and no damage of the crystal surface was observed during all experiments. Since the large aperture barium crystals can be grown, it can be assumed that THz output energy can be further increased by using higher energy laser pulses that will reveal the full potential of barium chalcogenides to be the compact sources of strong narrowband THz radiation.

In conclusion, generation of narrowband THz radiation in novel barium chalcogenides ($BaGa_4Se_7$, $BaGa_2GeSe_6$ and $BaGa_2GeS_6$) pumped by high power Cr:forsterite laser was reported for the first time. The highest optical-to-THz conversion efficiency of $0.8\times10^{-5}$ is reached in the Z-cut $BaGa_2GeS_6$ crystal in the saturation regime and corresponds to an output energy of 20 nJ. Pump fluence (intensity) of up to 60 mJ/cm$^2$ (600 GW/cm$^2$) was safely applied, and no damage of the crystal surface was observed during experiments. Lower values of conversion efficiency of the order of $5.4\times10^{-6}$ and $3.9\times10^{-6}$ were obtained in Z-cut BGGSe and BGSe selenide compounds, resulting in the THz energy of 13.5 and 7.5 nJ, respectively. Besides, polarization of generated THz pulses appeared to be linear, which is crucial for further applications of THz radiation. Thus, narrow-bandwidth THz pulses at the frequencies of 1.9 and 2.05 THz were obtained using selenides (BGSe and BGGSe), while for quaternary sulfide (BGGS) THz radiation generation with a relative bandwidth of 3.1% at 2.55 THz was observed. This study demonstrates the potential of novel quaternary Ba-compounds to be the compact sources of multi-cycle THz pulses.

## ACKNOWLEDGMENTS

This work was supported by Russian Science Foundation (RSF) (№25-72-00032). The equipment used in this work was purchased with the support of the Program for the Development of Moscow State University and National Project "Science and Universities". D.Z. Suleimanova is the scholar of the foundation for the advancement of theoretical physics and mathematics "BASIS".

## AUTHOR DECLARATIONS

**Conflict of Interest**

The authors have no conflicts to disclose.

**Author Contributions**

**D. Z. Suleimanova:** Formal Analysis, Investigation (equal), Software (equal), Writing/Original Draft Preparation (equal), **E. A. Migal**: Investigation (equal), Methodology (supporting), Software (equal), Writing/Original Draft Preparation (equal), **D. V. Badikov:** Resources, Writing/Review & Editing (equal), **F. V. Potemkin:** Conceptualization, Methodology (lead), Funding Acquisition, Project Administration, Supervision, Writing/Review & Editing (equal).

## DATA AVAILABILITY

The data that support the findings of this study are available from the corresponding author upon reasonable request.